# New Insights into Traffic Dynamics: A Weighted Probabilistic Cellular Automaton Model

X.L.Li[1], H.Kuang[1,2], T.Song[1], S.Q. Dai[1] and Z.P.Li[3]

[1] Shanghai Institute of Applied Mathematics and Mechanics, Shanghai University - Shanghai 200072, China

[2] College of Physics and Electronic Engineering, Guangxi Normal University - Guilin 541004, China

[3] School of Electronics and Information Engineering, Tongji University - Shanghai 201804, China

From the macroscopic viewpoint for describing the acceleration behavior of drivers, this letter presents a weighted probabilistic cellular automaton model (the WP model, for short) by introducing a kind of random acceleration probabilistic distribution function. The fundamental diagrams, the spatio-temporal pattern are analyzed in detail. It is shown that the presented model leads to the results consistent with the empirical data rather well, nonlinear velocity-density relationship exists in lower density region, and a new kind of traffic phenomenon called neo-synchronized flow is resulted. Furthermore, we give the criterion for distinguishing the high-speed and low-speed neo-synchronized flows and clarify the mechanism of this kind of traffic phenomena. In addition, the result that the time evolution of distribution of headways is displayed as a normal distribution further validates the reasonability of the neo-synchronized flow. These findings suggest that the diversity and randomicity of drivers and vehicles has indeed remarkable effect on traffic dynamics.

## 1. Introduction

Over the past decades, traffic problems have attracted considerable attention and various modeling approaches have been presented [1,2], such as car-following models, cellular automaton models, gas kinetic models and hydrodynamic models, etc. Here we will focus on cellular automaton (CA) models [1,3]. Compared with others, CA models are conceptually simpler, and can be easily implemented on computers for numerical investigations. There are two basic CA models describing singe-lane traffic flow: the Nagel-Schreckenberg (NaSch) model [4] and the Fukui-Ishibashi (FI) model [5]. Both of them are defined on a one-dimensional lattice consisting of $L$ sites with periodic boundary conditions. Each site is either occupied by a vehicle, or is empty. The velocity of each vehicle is an integer between zero and $V_{max}$. Let $x_i^t$ denote the position of the $i$ th car at time $t$, and $x_{i+1}^t$ the position of its preceding car at time $t$. Then the system evolves according to the synchronous rules, resulting in

$$x_i^{t+1} = x_i^t + v_i^{t+1} \tag{1}$$

where

$$v_i^{t+1} = \min\{v_{max}, v_i^t + 1, \Delta x_i\} \tag{2}$$

in the NaSch model and

$$v_i^{t+1} = \min\{v_{max}, \Delta x_i\} \tag{3}$$

in the FI model, in which $\Delta x_i$ is the headway of the $i$ th car with its preceding car. If we do not consider the randomization caused by other complicated influences, these two models differ only in acceleration rules, that is, the NaSch model restricts the cars to gradual acceleration, while the FI model allows for abrupt increase if there is enough empty spacing ahead. In both the update rules, the velocity of the $i$ th car depends on the headway. Based on the two models, many modified models have been proposed, among which are the VDR models [6], the $T^2$ model [7], the BJH model [8], and the VE model [9]. In addition, some related analytical and improved work about the FI model has been done by Wang et al [10,11] and Lee et al [12].

However, because the study of the CA traffic model has relatively short history, there have not yet been the 'best' CA traffic model which should be both realistic as well as simple. In this paper, we consider globally the different acceleration behaviors of drivers, which were not elucidated well in all previous work. In real traffic, there are various kinds of different drivers and vehicles on a road, and this diversity leads to different responses under the same traffic situation, e.g., aggressive drivers and pessimistic drivers (or fast vehicles and slow vehicles) have entirely different driving behaviors. In addition, such other stochastic factors as noise and external influences could affect the reactions to a certain extent. These reasons lead to the fact that not all drivers accelerate as those described in the FI model and NaSch model, where all drivers accelerate from zero to the possible maximum velocity or accelerate at the same probability under the same condition. Therefore, it is more realistic to take into account the diversity and randomicity of drivers and vehicle properties.

In view of the above analysis, we propose another new CA model from a totally different macroscopic viewpoint based on the probabilistic theory instead of using the traditional CA theory. Our basic idea is to introduce a kind of weighted random acceleration probability distribution function for random acceleration to synthetically describe the differences of drivers' own intentions, vehicle responses, noise and external influences. Here the introduction of this function indicates that the dependence of the driving strategy on the traffic state becomes essential. At the same time, it also implies the expectation of driving comfortably, i.e., avoiding both crashes and unnecessary large or small acceleration compared to the FI model and NaSch model.

## 2. Outline of the WP model

First of all, we explain the general framework of our model for single-lane traffic. $N$ vehicles are moving on a single-lane road divided into a one-dimensional array of $L$ sites. In this paper, parallel updating is adopted. Each site contains one vehicle at most, and collision and overtaking are thus prohibited (the so-called hard-core exclusion rule). Let $M_i^t$ be a stochastic variable which denotes the number of sites through which the $i$ th vehicle moves at time $t$. We introduce a probability distribution function $w_i^t(m)$ which gives the probability of the $i$ th vehicle hopping $m(m = 0,1,2,\cdots)$ sites ahead at time $t$, i.e., $M_i^t = m(m = 0,1,2,\cdots)$. The stochastic variable incorporates such uncertain effects as different driver and vehicle characteristics, incorrect driving operations, noise, and external influences. In this paper, we assume $m_{\max} = 5$. It should be noted that $m_{\max} = 5$ dose not imply the maximum velocities of every vehicle are all equal to 5 and we only use the probability density function to describe the mean effect of drivers, vehicles and external influence. The probability distribution function $w_i^t(m)$ is defined as

$$w_i^t(m) = \begin{cases} 1 & \Delta x_i = 0 \\ \dfrac{1}{\Delta x_i}(1-\dfrac{\alpha}{\gamma^{m+1}}) & m=0,1,2,\cdots,\Delta x_i-1; \Delta x_i \neq 0 \\ \dfrac{1}{\Delta x_i}(1-\dfrac{\beta}{\gamma^m}) & m=\Delta x_i; \Delta x_i \neq 0 \end{cases} \quad (4)$$

where $\alpha, \beta$ and $\gamma$ are underdetermined parameters which should meet the requirements that $\alpha, \beta, \gamma \in Z^+$ and $\alpha + \beta = \gamma$; $\Delta x_i = x_{i+1}^t - x_i^t - 1$ denotes the headway as before. Here it is assumed that if $\Delta x_i > 5$, $\Delta x_i = m_{max}$, i.e., the vehicle moves at the probability $w_i^t(m)$ of $\Delta x_i = 5$ with considering the speed limit on roads. It is necessary to point out that $w_i^t(m)$ has the following properties:

(1) $w_i^t(m)(m=0,1,2,\cdots,\Delta x_i)$ increases monotonically with increasing $m$ at given $\Delta x_i$, which means $w_i^t(m)/w_i^t(m-1) > 1$, for we know that drivers generally intend to move as fast as possible at constant $\Delta x_i$, but the ratio is close to 1 as $m$ increases, which reflects that the acceleration effect is almost the same at larger headway;

(2) $\sum_{m=0}^{\Delta x_i} w_i^t(m) = 1$, and $w_i^t(m) = 1$ for $\Delta x_i = 0$, which shows the randomization acceleration probability $w_i^t(m)$ obeys the unitary condition.

Based on the above analysis and realistic mathematical and physical consideration, the reasonable parameters are suggested as $\alpha = 2, \beta = 1, \gamma = 3$. The updating procedure consists of the following three steps.

- Determination of the randomization parameter $w_i^t(m)$ based on Eq.(4) at given $\Delta x_i$.

- Determination of the number of sites through which a vehicle passes (i.e., the probability velocity) according to the intention $w_i^t(m)$.

Based on Eq.(4), the set of probability density is composed of the following $\Delta x_i + 1$ subsets:

1). $[0, w_i^t(1))$;

2). $[w_i^t(1), w_i^t(1) + w_i^t(2))$;

...;

$\Delta x_i$). $[w_i^t(1)+\cdots+w_i^t(m-2), w_i^t(1)+\cdots+w_i^t(m-1))$;

$\Delta x_i +1$). $[w_i^t(1)+\cdots+w_i^t(m-1),1]$.

During the computation, as to a certain $\Delta x_i$, the random number $rand(\cdot)$ is produced by computer, and $rand(\cdot) \in [0,1]$. If $rand(\cdot)$ belongs to the $j$ th ($j=1,2,\cdots,\Delta x_i +1$) one of the above subsets, the number of sites which a vehicle hops is thus $m = j-1$.

$$v_i^t = m \tag{5}$$

- Updating of the new position of each vehicle

$$x_i^{t+1} = x_i^t + v_i^t (\forall i) \tag{6}$$

We call the probability distribution $w_i^t$ the intention because it is an intrinsic variable of drivers themselves. It brings uncertainty of operation into the traffic model and reflects a certain statistically mean effect.

## 3. Simulation results and discuss

The numerical simulation was performed according to the above updating rules under the periodic boundary condition. A one-dimensional lattice of $L$ sites and vehicles moving unidirectionally were considered. Each site was set to be 7.5m long, $L$ to be 1000, one time step to be 1s, being the order of the reaction time for humans, and the maximum velocity to be $v_{max} = 5$, corresponding to 135km/h. In order to characterize the behavior of the model, we determined the macroscopic quantities, including the global density $\rho$, the mean speed $V$, the mean flow $J$ defined as $\rho = N/L$, $V = \frac{1}{T}\sum_{t=t_0}^{t=T+t_0-1}\frac{1}{N}\sum_{n=1}^{N}v_n(t)$ and $J = \rho V$. In numerical simulation, the first $5 \times 10^4$ time-steps of each run was put away in order to remove the transient effects, and then the data were recorded in successive $1 \times 10^4$ time-steps. The mean velocity was obtained by averaging over 30 runs of simulations.

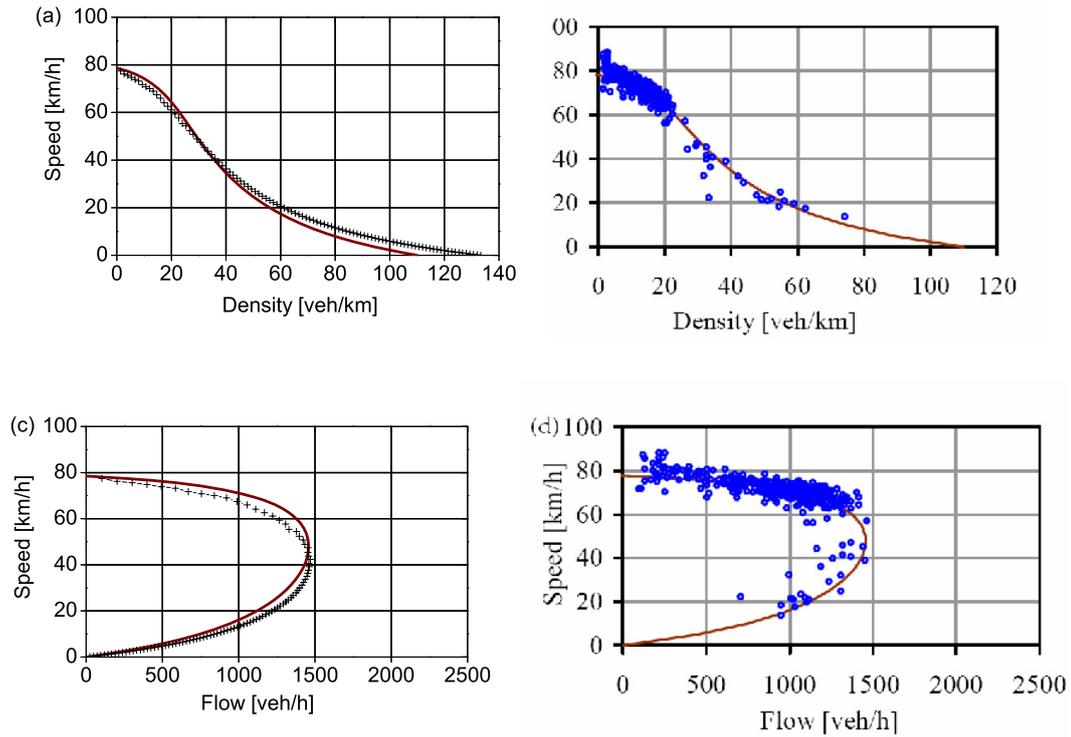

**Fig. 1. The comparison of results given with the WP model, and Van Aerde model, and the empirical data on Road 3 (Molduhraun - Arterial highway)presented in Ref.[13]. (a)The speed-density curves of the WP model and the Van Aerde model.(b)The speed-density curve of Van Aerde model and the empirical data. (c)The speed-flow curves of the WP model and the Van Aerde model.(d)The speed-flow curve of the WP model and the empirical data. In the figures, the red solid line denotes the Van Aerde model; the ++ stands for the WP model and the blue points are empirical data.**

Figure 1 presents the comparison of the results given with the WP model, the Van Aerde model and the empirical data presented in Ref.[13]. We can see the fundamental diagrams are in rather good agreement the empirical data. The results also validate the conclusion of general CA models well [14], i.e., on a single-lane road, the average velocity of vehicles is mainly determined by the slow vehicles. Meanwhile, the variation tendency of mean velocity is in good agreement with the empirical observations [15] instead of the occurrence of a plateau as appeared in other model CA models [3, 4], and a nonlinear flow-velocity-density relationship [16] exists.

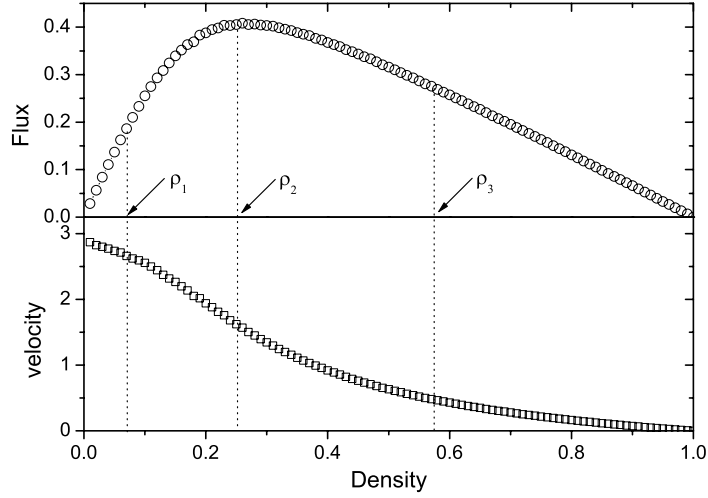

**Fig. 2. The fundamental diagram for the WP model: the upper section corresponds to the flux-density curve whereas the lower is the velocity-density curve.**

In order to analyze the traffic behavior revealed by the WP model, we further study the fundamental diagram (see Fig.2) and the spatio-temporal patterns (see Fig. 3). From the figures, we can find a new traffic state called neo-synchronized flow (the difference of it and the conventional synchronized flow will be expounded below). As the density is very low ($\rho < \rho_1$), the free flow exists. As the density exceeds $\rho_1$, the high-speed neo-synchronized flow begins to emerge (see Fig. 3(a)). As the density is in the range $\rho_1 < \rho < \rho_2$ and increases further, the neo-synchronized flow region gradually invades the free flow region and simultaneously, the region of the high-speed neo-synchronized flow is gradually amplified. As the density is equal to $\rho_2$, the flux reaches its maximum. In the range $\rho_1 < \rho < \rho_2$, the velocity of vehicles decreases from 2.6 to 1.5, whereas the flux increases from 0.17 to 0.41, which is the most basic phenomenon in the synchronized flow [17]. As the density is in the range $\rho_2 < \rho < \rho_3$, the high-speed neo-synchronized flow gradually transforms into the low-speed neo-synchronized flow (see Figs. 3(b) and (c)), in which the velocity is lower than that in the range $\rho_1 < \rho < \rho_2$ but the flux is still high, but a large area of jam can not appear. As the density exceeds $\rho_3$, both the velocity and flow are very low but the wide moving jam does not occur.

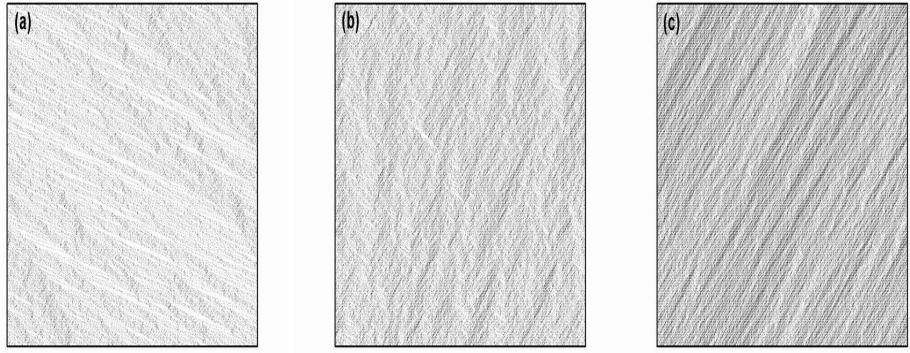

**Fig. 3. The spatio-temporal pattern of the WP model from an initial stochastic distribution. (a)** $\rho = 0.15$, **(b)** $\rho = 0.3$ **and (c)** $\rho = 0.5$. **The horizontal direction is space between** $200 \sim 700$ **cells and the vertical direction (downwards) is time in the period of** $49000 \sim 49600$.

The above analysis is mainly focused on the traffic phenomenon, and now turning to elucidate the mechanism of this kind of traffic phenomena. We first give a definition of the neo-synchronized flow, consisting of high-speed and low-speed neo-synchronized flows. Let us consider the velocity-density curve given in Fig. 2, and introduce the variables $\xi, \eta$ defined as

$$\xi = \frac{\Delta v}{\Delta \rho}; \eta = \frac{\Delta^2 v}{\Delta \rho^2} \tag{7}$$

where $\xi$ and $\eta$ are the first- and second-order difference quotients of velocity with respect to density, respectively. $\xi$ stands for the magnitude of change of velocity with respect to density and meets the condition: $\xi < 0$ and $\xi \to 0$ as $\rho \to 1$; and $\eta$ for the magnitude of change of $\xi$ with respect to density, which, in nature, discloses the degree of drivers' reflection to changing density and can be described as the "velocity fluctuation".

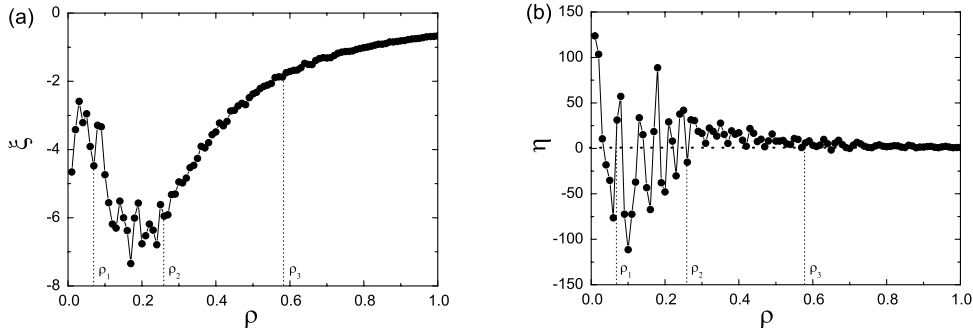

**4. The difference quotients of velocity corresponding to Fig. 2. (a)** $\xi = \frac{\Delta v}{\Delta \rho}$, **(b)** $\eta = \frac{\Delta^2 v}{\Delta \rho^2}$.

Figure 4 shows the difference quotients of velocity corresponding to Fig. 2 with the definition in Eq.(7). We can see clearly that: (1) as $\rho_1 < \rho < \rho_2$, $\xi$ decreases first and then increases a little for the density up to $\rho_2$. Besides, the fluctuation of $\xi$ and $\eta$ is larger and the difference quotients are unstable. These can be the criterion of high-speed neo-synchronized flow, and its velocity variation range is $56.7 \pm 13.5$ km/h; (2) as $\rho_2 < \rho < \rho_3$, $\xi$ is an increasing one, and the fluctuations of $\xi$ and $\eta$ are almost zero. These can be the criterion of low-speed neo-synchronized flow, and its velocity variation scope is $28 \pm 15$ km/h. These results reflect that in the high-speed neo-synchronized flow, the drivers can change vehicle velocities more easily with the variation of density because of the introduction of random probabilistic distribution function, where drivers have more choices at larger headway than at smaller headway, which causes the velocity to fluctuate intensively. This conclusion is similar to that in Refs.[18,19], where for synchronized flow, a certain range of velocity fluctuations of the vehicle should be acceptable. In addition, the random probabilistic distribution function could also explain why the spatio-temporal patterns qualitatively look the same and have some of the same characteristic features in Fig. 3, because there are always some drivers continuously have the tendency to almost collide with each other. As a consequence, the system is littered with mini-jams in both the low and high density regimes. This also reflects all spatio-temporal patterns in the system are of a chaotic nature (i.e., nonlinear with stochastic noise)[20]. These obtained results, to a certain extent, validate the fact that the appearance of synchronized flow can be reproduced by a mixture of different vehicle types assumption [21] as well as give a new insight into the explanation of traffic phenomena from the viewpoint of probability.

Note that the term about "light and heavy synchronized flow" had been found in Ref.[22], but with above specific analysis, we find some obvious differences between the synchronized flow in this paper and that in Ref.[22] as follows: (1) the scope of light and heavy synchronized flow is 24-60km/h in Ref.[22], while in our model the scope of neo-synchronized flow is extended to 13-70km/h and the acceptable velocity fluctuation of high-speed and low-speed neo-synchronized flow are also given; (2) the spatio-temporal patterns display no obvious large jam in the appearance of a neo-synchronized traffic flow, in which vehicles are all moving although they have a relatively lower speed; (3) spatio-temporal pattern in Fig. 3 is obtained from the initial stochastic distribution and the same spatio-temporal patterns can also be obtained from the initial homogeneous distribution and from an initial mega-jam distribution, which indicates the initial distribution has no influence on the patterns. Knowing from the synchronized flow in Ref.[22], we call this new phenomenon in this letter as the neo-synchronized flow.

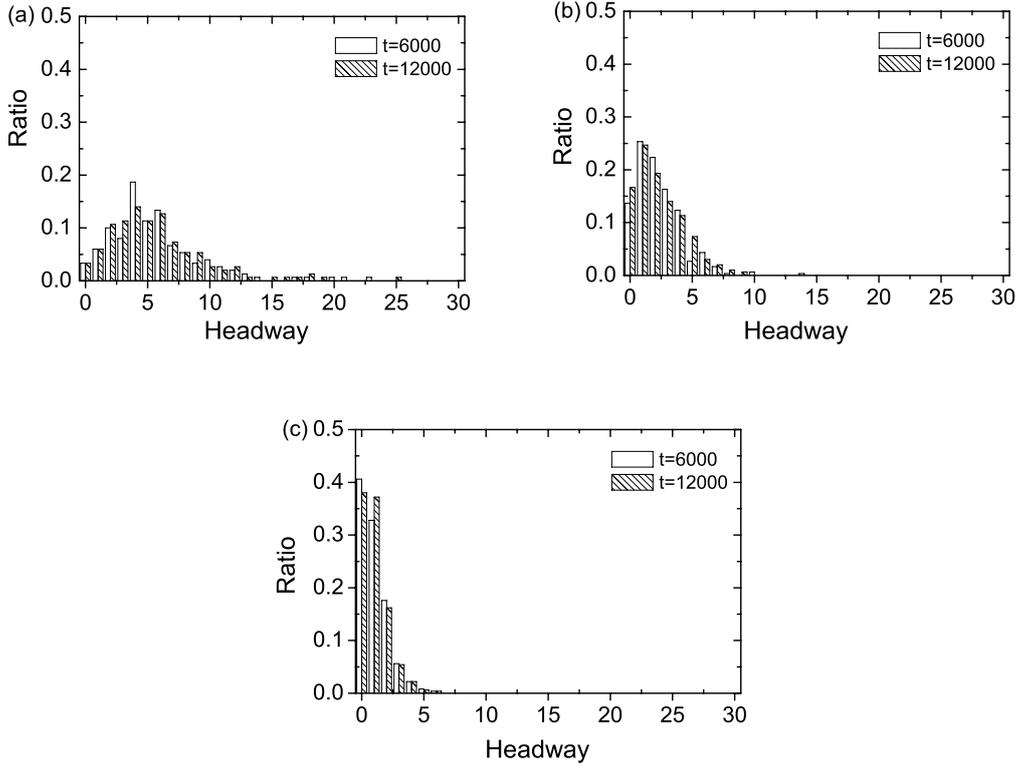

**Fig. 5.** The distribution of the headway with which the vehicles move at time step $t = 6000$ and $t = 12000$ of the WP model. **(a)** $\rho = 0.15$, **(b)** $\rho = 0.3$ and **(c)** $\rho = 0.5$.

Let us study the time evolution of distribution of headway. Figure 5 shows the distribution of headways at two time stages under different density $\rho = 0.15, 0.3, 0.5$, respectively. We find that the distribution of headways of the WP model changes significantly. In particular, the ratio of vehicles with null headway decrease evidently and the vehicles centralize at adjacent region $\frac{1}{\rho} - 1$ after a sufficiently time evolution, displayed as a normal distribution, which is consistent with the spatio-temporal patterns in Figure 3. In other words, on the one hand, this represents jams does not easily appears in this model, on the other hand, it shows the normal distribution of vehicles leads to the availability of road and flux increase. All these results further validate the reasonability of neo-synchronized flow.

## 4. Summary and Perspectives

In conclusion, our aim in this letter is to deliver a better understanding of the simulation of the traffic flow using a kind of totally different CA theory. To do so, beginning with a general scheme, we have proposed a new weighted probabilistic CA model, in which we have introduced a probability distribution function. The new model can be more adapted to reflecting some uncertain effects such as different driver and vehicle characteristics, driver error false operation and external influences. Through numerical simulations and comparison with the empirical data, it can be concluded that (i) the obtained fundamental diagrams are in good agreement with the empirical observations; (ii)the nonlinear velocity-density relationship is further validated; (iii)a new kind traffic behavior called neo-synchronized flow can be found; (iv)the criterion of neo-synchronized flow is rational and practicable; (v)time evolution of distribution of headway is displayed as a normal distribution; (vi)the basic idea of the new model is different from the conventional CA model and it has a simpler form and reasonable physical meaning.

Finally, the method, especially the modeling idea developed herein for all kinds of discrete dynamical system with statistical behavior, is actually more general. The similarities of some statistical properties of the investigated vehicle system, pedestrian flow system and granular system strongly suggest that a wide class of different phenomena can be described by the same type of modeling approach. The corresponding universality could complement the developed interdisciplinary research in the recent years. Further studies, e.g., extending the WP model and comparing with other CA models, will be given in subsequent publications.

This work was supported by the National Basic Research Program of China (Grant No.2006CB705500), the National Natural Science Foundation of China (Grant No. 10532060 and 10562001), the Special Research Fund for the Doctoral Program in Higher Education of China (Grant No.SRFDP 20040280014) and the Shanghai

Leading Academic Discipline Project (Grant No.Y0103).